\documentclass[%
 reprint,
superscriptaddress,
 amsmath,amssymb,
 aps,
 pre,
floatfix
]{revtex4-2}

\usepackage{graphicx}
\usepackage{dcolumn}
\usepackage{bm}
\usepackage[hidelinks]{hyperref}
\usepackage{xcolor}
\usepackage[normalem]{ulem}
\definecolor{ra}{rgb}{0.8, 0.0, 0.0}

\begin{document}

\preprint{APS/123-QED}
\title{Extending machine learning classification capabilities with histogram reweighting}
\author{Dimitrios Bachtis}
\email{dimitrios.bachtis@swansea.ac.uk}
\affiliation{Department of Mathematics,  Swansea University, Bay Campus, SA1 8EN, Swansea, Wales, UK}
\author{Gert Aarts}
\email{g.aarts@swansea.ac.uk}
\affiliation{Department of Physics, Swansea University, Singleton Campus, SA2 8PP, Swansea, Wales, UK}
\author{Biagio Lucini}
\email{b.lucini@swansea.ac.uk}
\affiliation{Department of Mathematics,  Swansea University, Bay Campus, SA1 8EN, Swansea, Wales, UK}%
\affiliation{Swansea Academy of Advanced Computing, Swansea University, Bay Campus, SA1 8EN, Swansea, Wales, UK}

\include{ms.bib}

\date{April 29, 2020}

\begin{abstract}

We propose the use of Monte Carlo histogram reweighting to extrapolate predictions of machine learning methods. In our approach, we treat the output from a convolutional neural network as an observable in a statistical system, enabling its extrapolation over continuous ranges in parameter space. We demonstrate our proposal using the phase transition in the two-dimensional Ising model. By interpreting the output of the neural network as an order parameter, we explore connections with known observables in the system and investigate its scaling behaviour. A finite size scaling analysis is conducted based on quantities derived from the neural network that yields accurate estimates for the critical exponents and the critical temperature. The method improves the prospects of acquiring precision measurements from machine learning in physical systems without an order parameter and those where direct sampling in regions of parameter space might not be possible.

\end{abstract}

\maketitle

\section{\label{sec:level1}Introduction}

Machine learning has recently emerged as an omnipresent tool across a vast number of research fields. A major milestone towards its wide success has been  the unprecedented capability of deep neural networks to automatically extract hierarchical structures in data \citep{GoodBengCour16}. Historical implementations of machine learning revolved around problems in image recognition and natural language processing but recent advances have encompassed the physical sciences \citep{Carleo_2019}. For a short review of machine learning for quantum matter see Ref.~\citep{2020arXiv200311040C}.

On the forefront of modern approaches, there have been significant contributions in the realm of computational physics.  Notably, machine learning was employed to study phase transitions in classical and quantum many-body systems \citep{Carrasquilla2017,vanNieuwenburg2017}. The aim is typically the separation of phases in a system by relying on supervised, unsupervised or semi-supervised learning of its configurations.  Neural networks \citep{PhysRevB.95.245134,Broecker2017,PhysRevX.7.031038,doi:10.7566/JPSJ.86.063001,PhysRevB.97.045207, 2017arXiv170700663B,PhysRevB.94.165134,PhysRevB.97.174435, 2019arXiv190303506A,PhysRevB.99.075418,PhysRevE.100.052312,PhysRevLett.121.245701,PhysRevB.98.060301,PhysRevLett.120.257204}, support vector machines \citep{PhysRevB.96.205146,GIANNETTI2019114639,PhysRevB.99.060404,PhysRevB.99.104410}, principal component analysis \citep{PhysRevB.94.195105,PhysRevE.95.062122, PhysRevB.96.144432,PhysRevB.96.195138, PhysRevB.99.054208} and a variety of algorithms \citep{PhysRevE.96.022140,PhysRevE.97.013306, Rodriguez-Nieva2019,PhysRevLett.121.255702,PhysRevE.99.023304}, have been implemented to achieve this goal. Within these approaches, the Ising model, due to its simplicity, analytical solution \citep{PhysRev.65.117}, and non-trivial phase structure, frequently acts as a prototypical testing ground to demonstrate results. 

In addition, efficient Monte Carlo sampling was realized through the construction of effective Hamiltonians in physical systems \citep*{PhysRevB.95.041101,PhysRevB.95.035105,PhysRevE.96.051301}. Among these approaches, the self-learning Monte Carlo method was extended to continuous-time, quantum and hybrid Monte Carlo \citep*{PhysRevB.96.161102,PhysRevB.96.041119,PhysRevB.97.205140,PhysRevB.98.041102,PhysRevB.100.020302,2019arXiv190902255N,PhysRevB.100.045153,PhysRevB.101.115111,PhysRevB.101.064308,PhysRevB.98.235145,PhysRevE.100.043301}. Deep reinforcement learning was utilized to generate ground states through the training of a machine learning agent, neural autoregressive models have been applied within variational settings, and Boltzmann generators have been introduced to produce unbiased equilibrium samples in condensed-matter and protein systems \citep{PhysRevE.99.062106,Noe,PhysRevLett.122.080602,PhysRevE.101.023304}. In lattice field theories generative and regressive neural networks have been implemented \citep{PhysRevD.100.011501}, and sampling with flow-based methods has led to reduced autocorrelation times \citep*{PhysRevD.100.034515,2020arXiv200306413K}.

\begin{figure*}
\includegraphics[width=16.2cm]{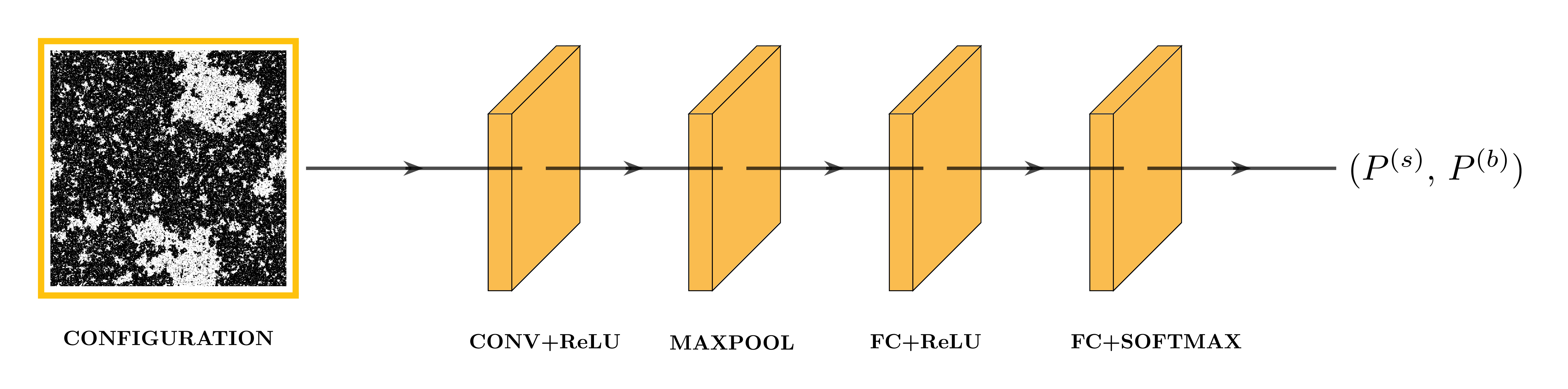}
\caption{\label{fig:conv2d}The architecture of the convolutional neural network (see App.~\ref{App:CNN}). Importance-sampled configurations are presented as input to the neural network and are further processed through a series of transformations. The values of the output vector denote the probability that a configuration belongs in an associated phase. The probabilities are used as observables to be reweighted.}
\end{figure*}

Along these lines, there is an increasing need for tools that improve computational efficiency and simultaneously enable the extraction of more information from available machine learning predictions. Traditionally, in statistical mechanics, such pursuits are achieved with the use of Monte Carlo histogram reweighting techniques \citep*{PhysRevLett.61.2635,PhysRevLett.63.1195}. It is then possible to acquire increased knowledge of observables by estimating them based on measurements from already conducted simulations. Furthermore, one can extrapolate in parameter space to glimpse into the behaviour of more complicated Hamiltonians. A proof of principle demonstration concerns reweighting from a zero external magnetic field to a non-zero value in the Ising model \citep{PhysRevLett.61.2635}.

In this article, we introduce histogram reweighting to supervised machine learning.  In particular, we explore if the output from a neural network in a classification problem can be treated as an observable in a statistical system and consequently be extrapolated to continuous ranges - providing a tool for further exploration without acquiring additional data and potentially even when direct sampling is not possible. We further interpret the output of the neural network as an effective order parameter within the context of a phase identification task, and propose reweighting as a means to explore connections with standard thermodynamic observables of the system under consideration. Finally, we search for scaling behaviour in quantities derived from the machine learning algorithm with an aim to study the infinite-volume limit of the statistical system. 

To establish the method, we apply it to the two-dimensional Ising model, a system that undergoes a second-order phase transition from a broken-symmetry to a symmetric phase. Using our developments, we present an accurate calculation of the critical point and the critical exponents by relying exclusively on quantities derived from the machine learning implementation and their reweighted extrapolations.

\section{\label{sec:level2}Histogram Reweighting}

We consider a generic statistical system described by an action (or Hamiltonian) $S=\sum_{k}g^{(k)} S^{(k)}$, which separates in terms of a set of parameters $\lbrace g^{(k)} \rbrace$. During a Markov chain Monte Carlo simulation of the system we sample a representative subset of states $\sigma_{1},\ldots,\sigma_{N}$ based on a Boltzmann probability distribution:
\begin{equation} \label{boltz}
p_{\sigma_i}= \frac{\exp[{-\sum_{k} g^{(k)} S_{\sigma_i}^{(k)}}]}{\sum_{\sigma} \exp[{-\sum_{k} g^{(k)} S_{\sigma}^{(k)}}]},
\end{equation}
where $Z=\sum_{\sigma} \exp[{-\sum_{k} g^{(k)} S_{\sigma}^{(k)}}]$ is the partition function and the sum is over all possible states $\sigma$  in the system.  The expectation value of an arbitrary observable $O$ is then given by:
\begin{equation} \label{estim}
\langle O \rangle=\frac{\sum_{i=1}^{N} {O_{\sigma_i} \tilde{p}_{\sigma_i}^{-1} \exp[{-\sum_{k} g^{(k)} S_{\sigma_i}^{(k)}}}]}{\sum_{i=1}^{N} \tilde{p}_{\sigma_i}^{-1}  \exp[{-\sum_{k} g^{(k)} S_{\sigma_i}^{(k)}}]},
\end{equation}
where $\tilde{p}_{\sigma_i}$ are the probabilities used to sample configurations from the equilibrium distribution. We now consider the probabilities for a set of parameter values $\lbrace g_{0}^{(k)} \rbrace$ that are sufficiently adjacent to $\lbrace g^{(k)} \rbrace$ in parameter space, given by:
\begin{equation}
{p}_{\sigma_i}^{(0)} = \frac{\exp[{-\sum_{k} g_{0}^{(k)} S_{\sigma_i}^{(k)}}]}{\sum_{\sigma} \exp[{-\sum_{k} g_{0}^{(k)} S_{\sigma}^{(k)}}]}.
\end{equation}

After substituting $\tilde{p}_{\sigma_i}$ with ${p}_{\sigma_i}^{(0)}$ in Eq.~(\ref{estim}), we arrive at reweighting equation:
\begin{equation} \label{sh}
\langle O \rangle_{\lbrace g^{(k)} \rbrace}=\frac{\sum_{i=1}^{N} {O_{\sigma_i}  \exp{[- \sum_{k} (g^{(k)}-g^{(k)}_{0})  S_{\sigma_i}^{(k)}}}]}{\sum_{i=1}^{N}   \exp{[- \sum_{k} (g^{(k)} - g^{(k)}_{0}) S_{{\sigma_i}}^{(k)}}]}.
\end{equation}

Given a series of Markov chain Monte Carlo measurements $O_{\sigma_i}$ for a set of parameters $\lbrace g_{0}^{(k)} \rbrace$, Eq.~(\ref{sh}) enables the calculation of expectation values for extrapolated sets of $\lbrace g^{(k)} \rbrace$.  Successful extrapolations of observables should lie within adjacent parameter ranges where the associated action histograms have markedly large values (e.g. see Ref.~\citep{newmanb99}).
\begin{figure*}
\includegraphics[width=8.6cm]{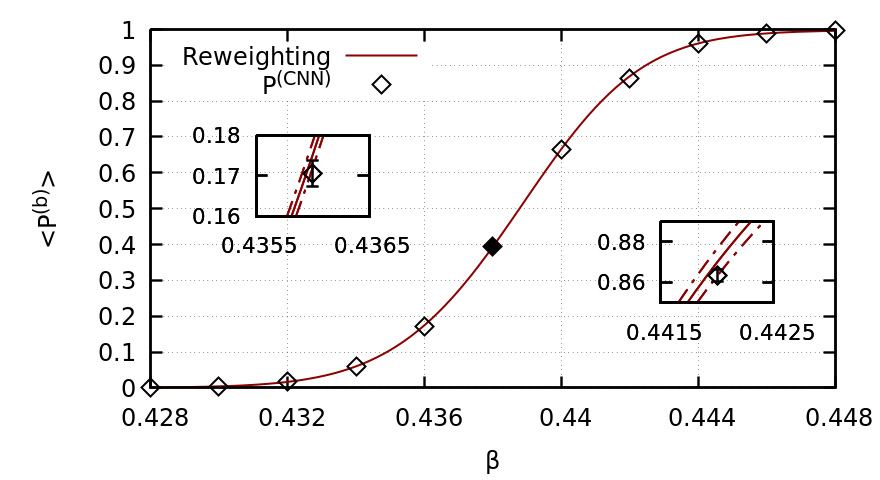} 
\includegraphics[width=8.6cm]{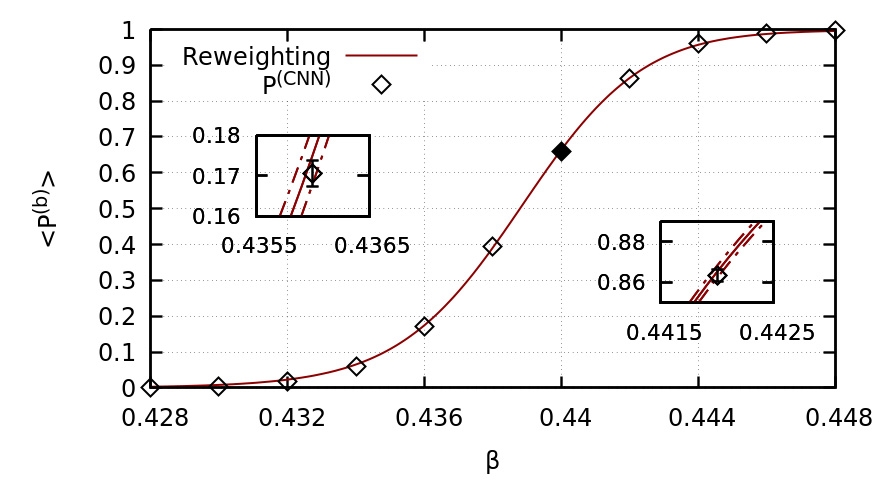}
\caption{\label{fig:sherr}Reweighting for the neural network output probability versus the inverse temperature of the 2D Ising model with lattice size $L=128$. The reweighted extrapolations are depicted by the line. The filled point corresponds to the parameter choice $\beta_{0}=0.438$ (left) and $\beta_{0}=0.44$ (right) used to conduct reweighting. Empty points are actual predictions from the neural network on Monte Carlo datasets, added for comparison. The dashed lines, only visible in the insets, indicate the statistical uncertainty. The training was conducted for $\beta \leq 0.41$ and $\beta \geq 0.47$. }
\end{figure*}

\section{\label{sec:level3}Reweighting of Machine Learning Output}

We employ reweighting techniques to study the phase transition in the two-dimensional Ising model (see App.~\ref{App:Ising}) by formulating the phase identification task as a classification problem.  We create the datasets using Markov chain Monte Carlo simulations with the Wolff algorithm \citep{PhysRevLett.62.361}. The training data are comprised of a set of 1000 uncorrelated configurations at each training point, with 100 configurations chosen as a cross-validation set. The range of inverse temperatures chosen to generate configurations used for training is $0.32,\ldots,0.41$ in the symmetric phase and $0.47,\ldots,0.56$ in the broken-symmetry phase with a step of $0.01$. The ranges are chosen to be distant from the critical inverse temperature $\beta_{c}\approx 0.440687$. We train the convolutional neural network (CNN) for lattice sizes $L=128,\ldots,760$. We implement the neural network architecture (see App.~\ref{App:CNN}) with TensorFlow and the Keras library \citep*{tensorflow2015-whitepaper,chollet2015keras}. The presence of a phase transition makes the convolutional neural network a well suited choice to learn spatial dependencies across configurations in different phases.

After training is completed, we present a configuration to the convolutional neural network to predict its associated classification label. The values of the output vector in the classification task sum up to one and are interpreted as the probability $P_{\sigma_i}$ that a configuration $\sigma_{i}$ belongs in the corresponding phase. When referring explicitly to the probabilities associated with the symmetric and the broken-symmetry phase we will denote them as $P_{\sigma_i}^{(s)}$ and $P_{\sigma_i}^{(b)}$, respectively, with $P_{\sigma_i}^{(s)}+P_{\sigma_i}^{(b)}=1$. 

In accordance with the sampling procedure which is carried through a Markov chain Monte Carlo simulation, each configuration appears in the chain of states as dictated by an associated Boltzmann weight. As depicted in Fig.~\ref{fig:conv2d},  the mathematical operation of convolution acts on an importance-sampled configuration and a series of additional transformations imposed by the neural network lead to the calculation of the probability $P_{\sigma_i}$. We therefore interpret $P$ as an observable of the system:
\begin{equation}
\langle P \rangle = \sum_{\sigma} P_{\sigma} p_{\sigma}= \frac{\sum_{\sigma}  P_{\sigma}  \exp[{-\sum_{k} g^{(k)} S_{\sigma}^{(k)}}]}{\sum_{\sigma} \exp[{-\sum_{k} g^{(k)} S_{\sigma}^{(k)}}]}. 
\end{equation}

In this framework, the probability $P$ can be extrapolated with histogram reweighting over wide ranges of parameter values $\lbrace g^{(k)} \rbrace$. Specifically for the case of the Ising model, reweighting in terms of inverse temperatures reduces Eq.~(\ref{sh}) to:
\begin{equation} \label{shising}
\langle P \rangle_{\beta} =\frac{\sum_{i=1}^{N} {P_{\sigma_i}   \exp{[-( \beta-\beta_{0})  E_{\sigma_i})}}]}{\sum_{i=1}^{N}   \exp{[-( \beta - \beta_{0}) E_{{\sigma_i}}]}}.
\end{equation}

 In Fig.~\ref{fig:sherr}, we show the expectation value $\langle P^{(b)}\rangle$ as a function of $\beta$ for the Ising model with lattice size $L=128$. The values over this large span of inverse temperatures, depicted by the line, have been obtained by exclusively extrapolating the probabilities from configurations of one Monte Carlo dataset. To demonstrate that reweighting is generally applicable we consider two cases where the Monte Carlo dataset is simulated at  $\beta_{0}=0.438$ in the symmetric phase or $\beta_{0}=0.44$ in the broken-symmetry phase. The reweighting results are compared with actual calculations of the average probability, which are obtained from predictions of the convolutional neural network on independent Monte Carlo datasets. The reweighted extrapolations overlap within statistical errors with the values from actual calculations (see App.~\ref{App:Boot}), demonstrating that the method is accurate. In addition we note, by comparing the results of the two cases, that the statistical errors of extrapolations increase with the distance from the reweighting point.
 
 We observe that the average probability resembles an order parameter, with values that are consistent with zero and one at different phases. In addition this effective order parameter has emerged by features learned on configurations for sets of inverse temperatures which lie beyond a fixed distance from the critical point $\beta_{c}$. The neural network has then fully reconstructed an effective order parameter based on incomplete information and representation of the studied system.

 \begin{figure}[t]
\includegraphics[width=8.6cm]{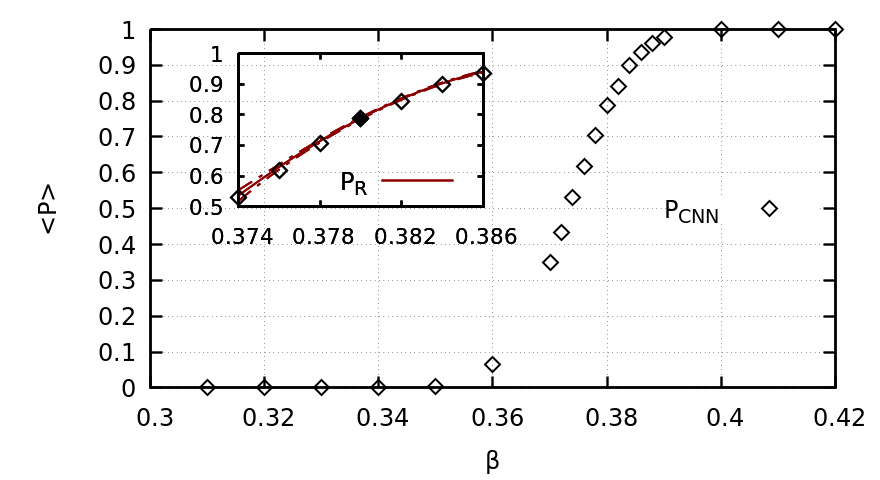}
\caption{\label{fig:other}Neural network output probability for the ensemble defined by configurations from inverse temperatures $\beta=0.41$ and $\beta=0.42$, versus inverse temperature $\beta$. Reweighting is depicted by the line in the inset, where statistical errors are visible. The filled point corresponds to the parameter choice $\beta=0.38$ used to conduct reweighting. Empty points are predictions of the CNN on independent Monte Carlo datasets, added for comparison. The training was conducted for $\beta=0.31, 0.32$ (labelled as 0) and $\beta=0.41, 0.42$ (labelled as 1).}
\end{figure}
 
We emphasize that the reweighting of machine learning devised observables is not inherently connected with the reconstruction of an effective order parameter in a statistical system and that it is generally applicable to learned neural network functions. As an example, we train a neural network to learn a function that acts as measure of the similarity between ensembles of configurations which reside exclusively within one phase of the system. One ensemble is comprised of configurations drawn from inverse temperatures $\beta=0.31$ and $\beta=0.32$, where the configurations are labeled as zero, and the second ensemble is comprised of configurations drawn from $\beta=0.41$ and $\beta=0.42$, labeled as one. 

 We apply the learned function to configurations of intermediate inverse temperatures to predict their associated label. The results are presented in Fig.~\ref{fig:other}, where the probability of a configuration belonging in the ensemble defined by $\beta=0.41$ and $\beta=0.42$ is depicted. Reweighting results are compared with calculations from independent Monte Carlo datasets, demonstrating that reweighting is accurate within statistical errors. The results evidence that reweighting is generally applicable to functions learned by a neural network in statistical systems.

\begin{figure}[t]
\includegraphics[width=8.6cm]{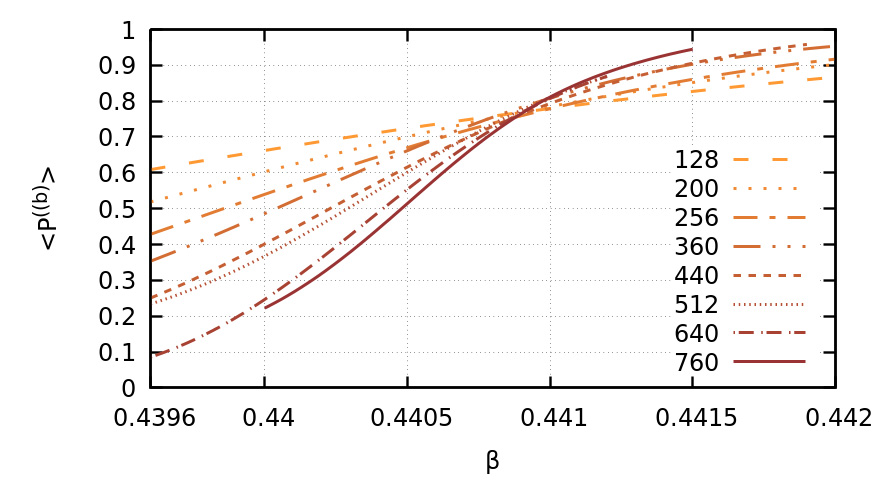}
\caption{\label{fig:magns}Reweighted extrapolations of the neural network output probability versus inverse temperature for various lattice sizes.}
\end{figure}

\section{\label{sec:level4}Finite Size Scaling on CNN-Derived Observables}

To further investigate the construction of an effective order parameter from a neural network trained for inverse temperatures $\beta \leq 0.41$ and $\beta \geq 0.43$ (see Fig.~\ref{fig:sherr}), we employ reweighting to draw the output probability $\langle P^{(b)} \rangle$ for a range of lattice sizes. Previous research (e.g.~in Refs.~\citep*{Carrasquilla2017,GIANNETTI2019114639}) has evidenced that decision making of a neural network seems to rely on some form of devised magnetization function. In Fig.~\ref{fig:magns} we note that the  probabilities for increasing lattice sizes become sharper near the critical point in a way that mimics the behaviour of the magnetization.  We recall that near a continuous phase transition and on a lattice of a finite size $L$, fluctuations such as the magnetic susceptibility $\chi$ have a maximum value.  A pseudo-critical point $\beta_{c}^{\chi}(L)$ is then associated with the maxima of the fluctuations which in the thermodynamic limit converges to the inverse critical temperature: 
\begin{equation}\label{eq:limit}
\lim_{L \to \infty} \beta_{c}^{\chi}= \beta_{c}.
\end{equation}

Considering that the neural network output probability manifests behaviour which is reminiscent to that of an effective order parameter, we proceed by investigating its fluctuations, weighed by the inverse temperature, which are defined as:
\begin{equation} \label{eq:probsusceq}
 \delta P = \beta V(\langle  P^{2} \rangle - \langle P \rangle^{2}).
\end{equation}

\begin{figure}[t]
\includegraphics[width=8.6cm]{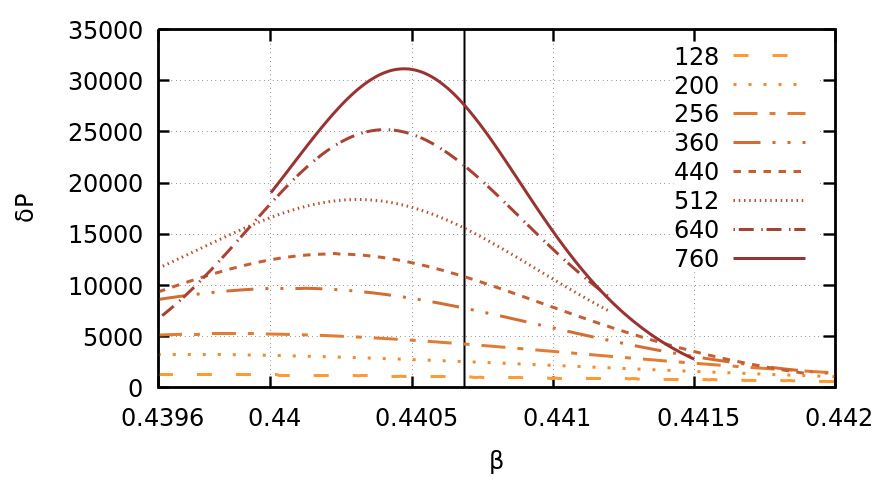}
\caption{\label{fig:susc}Fluctuations of the neural network output probability, Eq.~(\ref{eq:probsusceq}), for various lattice sizes. The vertical line corresponds to the known critical temperature $\beta_{c} \approx 0.440687 $ of the two-dimensional Ising model.}
\end{figure}

Since reweighting has been formulated in terms of an arbitrary observable, we use Eq.~(\ref{sh}) to estimate the expectation values of both observables $\langle P^{2} \rangle$, $\langle P \rangle$ and hence calculate the fluctuations of the probability which are depicted in Fig.~\ref{fig:susc}, without including statistical errors. We note that the inverse temperatures where the maximum values of the fluctuations $\delta P$ are located evidence a scaling behaviour with increasing lattice sizes.  As discussed above, such scaling behaviour of the fluctuations is anticipated for a quantity that acts as an effective order parameter, and it tentatively indicates a convergence towards the known inverse critical temperature of the Ising model $\beta_{c} \approx 0.440687$, which is depicted by the vertical line. We therefore associate pseudo-critical points $\beta_{c}^{P}(L)$ for the values of the maxima $\delta P_{max}$ to investigate their convergence in the thermodynamic limit (see Eq.~(\ref{eq:limit})), and to calculate multiple critical exponents in our subsequent quantitative analysis.

In order to estimate the correlation length exponent $\nu$ and the inverse critical temperature $\beta_{c}$ we note that, due to the divergence of the correlation length in the pseudo-critical region, the reduced temperature can be expressed as:
\begin{equation} \label{fit1}
|t|= \Big| \frac{\beta_{c}-\beta_{c}(L)}{\beta_{c}} \Big| \sim \xi^{-\frac{1}{\nu}} \sim L^{-\frac{1}{\nu}}.
\end{equation}

Consequently, without presuming any knowledge about the values of the inverse critical temperature and the correlation length exponent, we can calculate them simultaneously using Eq.~(\ref{fit1}).

In addition, we investigate if the fluctuations of the neural network output probability, which resembles an effective order parameter, are governed by the same critical exponent as the fluctuations of the conventional order parameter, which is the magnetization. We therefore perform a calculation for the magnetic susceptibility exponent $\gamma$ using the maximum values of the probability fluctuations:
\begin{equation} \label{fit2}
\delta P \sim L^{\frac{\gamma}{\nu}}.
\end{equation}

\begin{figure}[t]
\includegraphics[width=8.6cm]{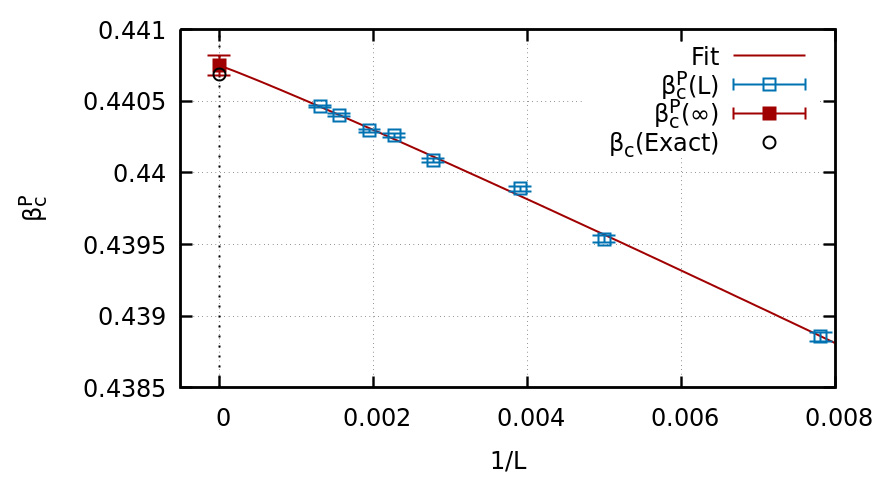}
\caption{\label{fig:fss1}Inverse pseudo-critical temperature versus inverse lattice size.}
\end{figure}

\begin{figure}[t]
\includegraphics[width=8.6cm]{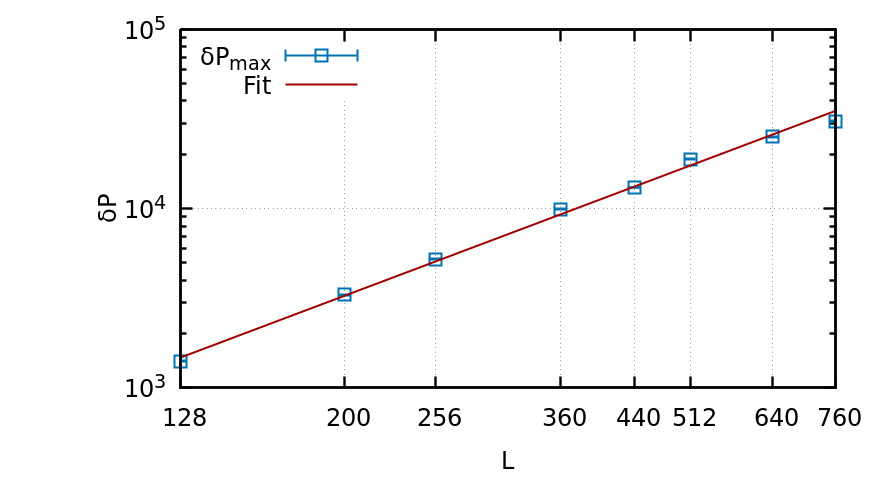}
\caption{\label{fig:fss2}Probability fluctuations versus lattice size on double logarithmic scale.}
\end{figure}

As visible in Figs.~\ref{fig:fss1} and \ref{fig:fss2}, we fit the data (see Tab.~\ref{tab:data}) for the pseudo-critical points and the maxima of the probability fluctuations using Eqs.~(\ref{fit1}) and (\ref{fit2}), respectively. The results of the finite size scaling analysis are given in Tab.~\ref{tab:table2}. We note that the obtained estimates for the critical exponents ($\nu=0.95(9)$, $\gamma/\nu=1.78(4)$) and the inverse critical temperature ($\beta_{c}=0.440749(68)$) of the Ising model are within statistical errors of the known values from Onsager's analytical solution ($\nu=1$, $\gamma/\nu=7/4$, $\beta_{c}=\ln(1+\sqrt{2})/2$). In the error analysis only statistical errors from predictions of the neural network on a finite Monte Carlo dataset were considered. 

\begin{table}[b]
\caption{\label{tab:data}
Pseudo-critical points $\beta_{c}^{P}(L)$ and maxima of the probability fluctuations $\delta P_{max}$ for various lattice sizes $L$ of the Ising model.}
\begin{ruledtabular}
\begin{tabular}{ccc}
$L$ &$\beta_{c}^{P}(L)$&$\delta P_{max}$ \\
\hline
128 & 0.438857(33) & 1409(6) \\
200 & 0.439536(24) & 3308(14) \\
256 & 0.439889(18) & 5233(24) \\
360 & 0.440088(13) & 9910(49) \\
440 & 0.440261(12) & 13138(71) \\
512 & 0.440292(10) & 18912(99) \\
640 & 0.440403(10) & 25215(218) \\
760 & 0.440465(8) & 30841(206) 
\end{tabular}
\end{ruledtabular}
\end{table}

\section{\label{sec:level5}Conclusions}

In this article we introduced histogram reweighting to supervised machine learning. By treating the output of a neural network in a phase identification task as an observable in the two-dimensional Ising model, we utilized reweighting to extrapolate it over continuous ranges in parameter space. We further interpreted the output as an effective order parameter and investigated its scaling behaviour. This resulted in a calculation of the correlation length and magnetic susceptibility critical exponents, as well as the inverse critical temperature, based on a finite size scaling analysis conducted on quantities derived from the neural network.
\begin{table}[b]
\caption{\label{tab:table2}
Critical exponents $\nu$,$\gamma/\nu$ and critical inverse temperature $\beta_{c}$ of the Ising model acquired by reweighting quantities of the neural network implementation and comparison with exact values from Onsager's analytical solution.}
\begin{ruledtabular}
\begin{tabular}{ccccc}
 &$\beta_{c}$&$\nu$ &$\gamma/\nu$ \\
\hline
CNN+Reweighting & 0.440749(68) & 0.95(9) & 1.78(4) \\
  Exact & $\ln(1+\sqrt{2})/2$ & $1$ & $7/4$  \\
&  $ \approx0.440687$ & & =1.75
\end{tabular}
\end{ruledtabular}
\end{table}

The extension of histogram reweighting to neural networks enables quantitative studies of phase transitions based on a synergistic relation between machine learning and statistical mechanics. Generalizing to multiple histogram reweighting is straightforward. The effective order parameter learned by the neural network on the spatial structure of configurations, where no explicit information about the symmetries of the Hamiltonian is introduced, can be studied with high precision using reweighting and could prove useful when a conventional order parameter is absent or unknown \citep{Carrasquilla2017}. Examples are phenomena that are currently under active investigation, such as topological superconductivity \citep{Sato_2017}, and the finite-temperature phase transition in quantum chromodynamics \citep{bors2010,PhysRevD.85.054503}. Finally, through multi-parameter reweighting, one could explore the extrapolation of machine learning predictions in regions of parameter space where direct sampling with Monte Carlo might not be possible. Such cases potentially include systems with a numerical sign problem \citep{Aarts_2016}.

\section{\label{sec:level6}Acknowledgements}

The authors received funding from the European Research Council (ERC) under the European Union's Horizon 2020 research and innovation programme under grant agreement No 813942. The work of GA and BL has been supported in part by the STFC Consolidated Grant ST/P00055X/1. The work of BL is further supported in part by the Royal Society Wolfson Research Merit Award WM170010. Numerical simulations have been performed on the Swansea SUNBIRD system. This  system is part of the Supercomputing Wales project, which is part-funded by the European Regional Development Fund (ERDF) via Welsh Government. We thank COST Action CA15213 THOR for support.

\appendix

\section{\label{App:Ising}Ising Model}

We consider the Ising model on a hypercubic two-dimensional square lattice with Hamiltonian:
\begin{equation}
E=-J \sum_{\langle ij \rangle} s_{i} s_{j} - h \sum_{i} s_{i},
\end{equation}
where $\langle ij \rangle$ denotes a sum over nearest neighbors, $J$ is the coupling constant which is set to one, and $h$ the external magnetic field which is set to zero. 

The system is invariant under a reflection symmetry $\lbrace s_{i} \rbrace \rightarrow \lbrace -s_{i} \rbrace$ that can be spontaneously broken. We define in the vicinity of a continuous phase transition, a dimensionless parameter called the reduced inverse temperature:
\begin{equation}
t= \frac{\beta_{c}-\beta}{\beta_{c}},
\end{equation}
where $\beta_{c}$ is the critical temperature. The divergence of the correlation length for a system in the thermodynamic limit $\xi=\xi(\beta,L=\infty)$  is given by:
\begin{equation}
\xi \sim |t|^{-\nu},
\end{equation}
with $\nu$ the correlation length critical exponent. Another observable of interest is the normalized magnetization:
\begin{equation}
 m =\frac{1}{V}\bigg| \sum_{i} s_{i} \bigg|, 
\end{equation}
where $V$ is the volume of the system. The magnetic susceptibility is then defined as the fluctuations of the magnetization:
\begin{equation}
 \chi= \beta V (\langle m^{2} \rangle - \langle m \rangle^{2} ),
\end{equation}
and has an associated critical exponent $\gamma$, defined via:
\begin{equation}
\chi \sim |t|^{-\gamma}.
\end{equation}
The sets of critical exponents determine a universality class.  Knowledge of two exponents is sufficient for the calculation of the remaining ones through the use of scaling relations (e.g. see Ref.~\cite{newmanb99}). The exact results for the two-dimensional Ising model are:
\begin{eqnarray}
&\nu =1,\\
&\gamma/\nu =7/4,\\
&\beta_{c} = \frac{1}{2} \ln(1+\sqrt{2}) \approx 0.440687.
\label{eq:one}
\end{eqnarray}

\section{\label{App:CNN}CNN Architecture}

The neural network architecture (see Fig.~\ref{fig:conv2d}) consists of a 2D convolutional layer with 64 filters of size $2 \times 2$ and a stride of 2, supplemented with a rectified linear unit (ReLU) activation function. The result is then passed to a $2 \times 2$ max-pooling layer and subsequently to a fully connected layer with 64 ReLU units. The output layer consists of two units with a softmax activation function, with values between $[0,1]$. Configurations in the symmetric phase are labeled as $(1,0)$ and in the broken-symmetry phase as $(0,1)$. We train the CNN until convergence using the Adam algorithm and a mini-batch size of 12. To speed up the learning in small lattices of $L \leq 256$, we choose a learning rate of $10^{-4}$ and reduce it by a factor of 10 for the remaining sizes. 

The architecture is selected using an empirical approach. Initially, the CNN is trained for lattice size $L = 128$ where its training and validation loss, as well as their difference, are monitored to be minimal. This certifies that the CNN accurately separates phases of the 2D Ising model based on the presented training data and can generalize well to unseen data. The approach is then extended successfully to larger lattice sizes, up to $L = 760$, covering the complete range of sizes used for the finite size scaling with the same CNN architecture. Architectures for more complicated systems can be derived using the same approach. When testing a different architecture, an increase or decrease in the number of trainable parameters might lead to overfitting or underfitting of data. The occurrence of underfitting or overfitting can be monitored based on the effect on the values of the validation loss.

\section{\label{App:Boot}Bootstrap Analysis}

The calculation of errors has been conducted with a boostrap analysis \citep{newmanb99}. This enables the elimination of any potential bias associated with the finite Monte Carlo generated sample. In particular, each Monte Carlo dataset, comprised of uncorrelated configurations, has been resampled a 1000 times for each lattice size. For each resampled dataset, reweighted extrapolations of the output probability and its fluctuations are acquired in a wide range of temperatures. The error for the extrapolated probability $P$ at each inverse temperature $\beta$ is given by equation:
\begin{equation}
\sigma = \sqrt{\overline{P^{2}}-\overline{P}^{2}},
\end{equation}
where the averages are performed over the bootstrap replicas. 

\bibliography{ms}
\end{document}